\documentclass[structabstract]{aa}  

\usepackage{enumerate}
\usepackage{graphicx}
\usepackage{txfonts}
\usepackage{natbib}

\newcommand{\beq}{ \begin{eqnarray} }
\newcommand{\eeq}{ \end{eqnarray} }

\newcommand{\boldnabla}{\mbox{\boldmath$\nabla$}}

\begin{document}
  
  \title{MHD Dynamical Relaxation of Coronal Magnetic Fields}

  \subtitle{I. Parallel Untwisted Magnetic Fields in 2D}

  \author{Jorge Fuentes-Fern\'andez, Clare E. Parnell and Alan W. Hood}

  \institute{School of Mathematics and Statistics, University of St Andrews, North Haugh, St Andrews, Fife, KY16 9SS, Scotland}
  
  \abstract
  {For the last thirty years, most of the studies on the relaxation of stressed magnetic fields in the solar environment have only considered the Lorentz force, neglecting plasma contributions, and therefore, limiting every equilibrium to that of a force-free field.}
  {Here we begin a study of the non-resistive evolution of finite beta plasmas and their relaxation to magnetohydrostatic states, where magnetic forces are balanced by plasma-pressure gradients, by using a simple 2D scenario involving a hydromagnetic disturbance to a uniform magnetic field. The final equilibrium state is predicted as a function of the initial disturbances, with aims to demonstrate what happens to the plasma during the relaxation process and to see what effects it has on the final equilibrium state.}
  {A set of numerical experiments are run using a full MHD code, with the relaxation driven by magnetoacoustic waves damped by viscous effects. The numerical results are compared with analytical calculations made within the linear regime, in which the whole process must remain adiabatic. Particular attention is paid to the thermodynamic behaviour of the plasma during the relaxation.}
  {The analytical predictions for the final non force-free equilibrium depend only on the initial perturbations and the total pressure of the system. It is found that these predictions hold surprisingly well even for amplitudes of the perturbation far outside the linear regime.}
  {Including the effects of a finite plasma beta in relaxation experiments leads to significant differences from the force-free case.}


  \maketitle


\section{Introduction}

The magnetic field of the solar corona is believed to evolve through a series of force-free states \citep{Heyvaerts84}. Since the solar corona involves a low-beta plasma in which magnetic forces dominate over plasma forces, this is not an unreasonable assumption, and so, most of the recent studies on the relaxation of coronal magnetic fields \citep[e.g.][]{Mackay06,Ji07,Inoue08,JanseLow09,Milleretal09,Pontin09} have been done by considering the approximation of an extremely tenuous plasma, for which the plasma pressure does not play an important role, and the persistent hydromagnetic structures of the solar corona are assumed to be in magnetic balance, with zero pressure gradients. On the other hand, only within the past few years, \citet{Ruan08} and \citet{Gary09} have started to consider the reconstruction of the global coronal magnetic field including a finite Lorentz force balanced by magnetic and gravity forces.

In addition, there are many codes available to calculate those force-free fields from the observed magnetic field in the photosphere \citep{Amari98,Wiegelmann06,Wiegelmann08,Schrijver06,Metcalf08}. These codes have been used with varying degrees of success to determine the magnetic field of solar flares and active regions \citep[e.g.][]{DeRosa09,Schrijver08,Regnier08,Wheatland09}. However, problems remain with these approaches. In particular, a non-linear force-free field determined from a line-of-sight photospheric magnetic field is not unique, but is one of an infinite number of possible solutions. This fact is well known and has been discussed by several authors \citep[see][]{Low06}.

Also, the low beta plasma assumption is only valid for the solar corona, but is not valid in the photosphere, chromosphere or much of the transition region where plasma effects become more important. Furthermore, in the solar corona, the tenuous plasma will be able to create a high beta in the surroundings of a magnetic null point \citep{McLaughlin06}. Moreover, even where the plasma has a low beta, there are still some effects on the final equilibrium of the magnetic field that will lead to energetic consequences as the field is relaxing to its new equilibrium state.

In particular, relaxation can involve the magnetic field evolving from a stressed state to a force-free field with a lower energy, and for a consistent scenario where the energy cannot escape the system, conservation of energy implies that the losses of magnetic energy must be converted into something else. \citet{Browning08} and \citet{Hood09} investigated Taylor relaxation \citep{Taylor74} through a series of non-linear 3D simulations, initiated by an ideal MHD instability. Although the initial state was force-free, the final state involved a high temperature plasma with a significant value for the plasma beta. Placing aside some extra contributions such as radiative losses, if a substantial fraction of the magnetic energy released goes into the internal energy, then the plasma beta cannot be small. Hence, considering the behaviour of the plasma is important even if it has little effect on the final magnetic equilibrium. On the other hand, \citet{Gary01} suggested the possibility that there is high beta plasma in the solar corona above active regions.

In this paper, we will start by considering a simple scenario, with a uniform 2D magnetic field. There will be no current sheet formation and no changes in the magnetic topology. The aim of the paper is to investigate the relaxation of the hydromagnetic fluid for various different values of the plasma beta. To perturb the system, we introduce a local small enhancement (or deficit) in the plasma pressure (or in the magnetic field), under the frozen-in condition. The subsequent relaxation leads to a new magnetohydrodynamic equilibrium in which the perturbation has been dissipated by viscous forces acting on the flow.

It is worthwhile mentioning that relaxation via Ohmic dissipation, due to the effect of resistivity, or magnetic diffusivity, represents a substantially different problem; While viscosity dissipates the plasma velocity, diffusivity tends to eliminate the electric current density, and such a relaxed state can only involve potential fields, which are mathematically well defined and are uniquely determined by the components of the magnetic field normal to the boundaries. Furthermore, the time-scales for an Ohmic relaxation in very high magnetic Reynolds number environments, such as the solar corona, are in general probably larger than the age of the Sun itself, outwith regions with very small length scales \citep[see][]{Priest82}.

By considering the effects of the plasma pressure in the relaxation, we are facing a totally different problem from that of the force-free relaxation studied by many others. In our non-resistive MHD relaxation, the plasma displacements driven by the initial pressure enhancement will carry the magnetic field with them, generating an electric current and a magnetic pressure. Hence, the resulting equilibrium will have to involve a balance between the Lorentz force and the plasma-pressure gradient.

The effects of including a finite plasma beta are relevant not only in the high plasma beta regions of the solar atmosphere such as the photosphere and chromosphere, but will also be relevant in the solar corona. Obvious regions where the plasma beta is likely to have a significant effect are in the vicinities of magnetic neutral points, where the magnetic field vanishes. These configurations will be the subject of a further paper, while in the present paper we study the plasma beta effects on the simplest magnetic configuration, a uniform magnetic field, which is absolutely general and might be compared with different solar environments such as a region in a coronal prominence or part of a coronal loop.

In Section \ref{sec:analytic}, we first present the setup of the two-dimensional linear problem. In Sections \ref{sec:1Dx} and \ref{sec:1Dy}, we solve the equations for a one-dimensional vertical and horizontal perturbation, respectively, and in \ref{sec:2D} we obtain the whole solution from a general two-dimensional linear perturbation, producing a practical and precise analytical solution to the problem. In Section \ref{sec:numexp}, we present a set of numerical experiments run using the Lare2D code \citep{Arber01}, and we compare these with the previous analytical results. The effects of the non-linear terms are considered when the magnitude of the perturbation is increased, in Section \ref{sec:nonlin}, followed by the summary and some conclusions which are presented in Section \ref{sec:summary}.


\section{Linear 2D Equations}\label{sec:analytic}

The initial set up involves a uniform magnetic field pointing in the vertical $y$-direction, ${\bf B}_0=B_0\hat{\bf e}_y$, and a background plasma of a constant gas pressure, density and temperature. The initial disturbances are supposed to be small, in order to stay in the linear regime. Expressing each quantity $q(x,y,t)$ as the sum of a background constant value plus a perturbation, $q(x,y,t)=q_0+q_1(x,y,t)$, and substituting those expressions into the viscous, non-resistive, two-dimensional MHD equations, with no gravity force for simplicity, and neglecting terms involving products of perturbations, we obtain the following set of linearized equations, in scalar form:
\begin{small}
\begin{eqnarray}
\frac{\partial \rho_1}{\partial t}&=&-\rho_0\boldnabla\cdot{\bf v}_1\;, \label{eq_rho1}\\
\rho_0\frac{\partial v_x}{\partial t}&=&-\frac{\partial p_1}{\partial x}-\frac{B_0}{\mu_0}\frac{\partial B_y}{\partial x}+\frac{B_0}{\mu_0}\frac{\partial B_x}{\partial y}+\rho_0\nu\left(\nabla^2 v_x+\frac{1}{3}\frac{\partial}{\partial x}(\boldnabla\cdot{\bf v}_1)\right)\;, \label{eq_vx}\\
\rho_0\frac{\partial v_y}{\partial t}&=&-\frac{\partial p_1}{\partial y}+\rho_0\nu\left(\nabla^2 v_y+\frac{1}{3}\frac{\partial}{\partial y}(\boldnabla\cdot{\bf v}_1)\right) \label{eq_vy}\; ,\\
\frac{\partial p_1}{\partial t}&=&-\gamma p_0 \boldnabla\cdot{\bf v}_1\;, \label{eq_p1}\\
\frac{\partial B_x}{\partial t}&=&B_0 \frac{\partial v_x}{\partial y}\;, \label{eq_Bx}\\
\frac{\partial B_y}{\partial t}&=&-B_0 \frac{\partial v_x}{\partial x}\;, \label{eq_By}
\end{eqnarray}
\end{small}
where $\rho_0$, $p_0$ and $B_0$ are the constant background plasma density, plasma pressure and magnetic field, $\nu$ is the kinematic viscosity, $\rho_1$, $p_1$ and ${\bf v}_1$ are the perturbations on plasma density, plasma pressure and velocity, and $v_x$, $v_y$, $B_x$ and $B_y$ are the $x$ and $y$ components of the perturbed velocity and perturbed magnetic field, respectively. Also, the plasma pressure, $p$, density, $\rho$, and internal energy, $\epsilon$, are related by the perfect gas law, for an ideal polytropic gas,
\begin{equation}
p=\rho\epsilon(\gamma-1)\;, \label{eq_gas}
\end{equation}
where $\gamma$ is the ratio of specific heats, often assumed to be $5/3$ for a highly ionised hydrogen plasma.

In a general scheme, viscosity would add a heating term to the energy equation, but this term is second order, and so, the process is adiabatic within the linear regime, and there is no heating of any kind taking place. Hence, the entropy per unit mass, $S=p/\rho^\gamma$, is conserved, for each single fluid element, and for the entire box.

From the conservation of entropy, a relation between the plasma pressure and density perturbations may be obtained, within first order (i.e. neglecting the terms involving products of perturbations):
\begin{eqnarray*}
\frac{p_0+p_1}{(\rho_0+\rho_1)^{\gamma}}&=&\frac{p_0}{\rho_0^{\gamma}}\left(1+\frac{p_1}{p_0}-\frac{\gamma\rho_1}{\rho_0}\right)\;=\;{\rm constant}\;.
\end{eqnarray*}
Hence,
\begin{eqnarray*}
\frac{\Delta p_1}{p_0}-\frac{\gamma\Delta\rho_1}{\rho_0}=0\;,
\end{eqnarray*}
where $\Delta$ indicates the difference between final and initial state of the perturbation, such that
\begin{equation}
\Delta p_1 = c^2_s \Delta\rho_1 \;, \label{adiabatic}
\end{equation}
where $c_s=\sqrt{\gamma p_0/\rho_0}$ is the sound speed.

To solve these equations, we first determine the solution for a one-dimensional perturbation, which depends only on $x$ (Sec. \ref{sec:1Dx}), then only on $y$ (Sec. \ref{sec:1Dy}), and finally, using the 1D results, we derive the solution for a more general 2D perturbation (Sec. \ref{sec:2D}).


\subsection{Perturbation dependent on x} \label{sec:1Dx}

Consider a perturbation varying only in the direction perpendicular to the magnetic field lines, $x$. The magnetic field vector will have a non-zero $y$-component, ${\bf B}_1(x,t)=B_y(x,t)\hat{\bf e}_y$, while the velocity will have a non-zero $x$-component, ${\bf v}_1(x,t)=v_x(x,t)\hat{\bf e}_x$. Equations (\ref{eq_rho1}) to (\ref{eq_By}) reduce to
\begin{eqnarray}
\frac{\partial \rho_1}{\partial t}&=&-\rho_0\frac{\partial v_x}{\partial x}\;, \label{eq_rho1Dx}\\
\rho_0\frac{\partial v_x}{\partial t}&=&-\frac{\partial p_T}{\partial x}+\rho_0\nu^{\prime}\frac{\partial^2 v_x}{\partial x^2}\;, \label{eq_v1Dx}\\
\frac{\partial p_1}{\partial t}&=&-\gamma p_0 \frac{\partial v_x}{\partial x}\;, \label{eq_p1Dx} \\
\frac{\partial B_y}{\partial t}&=&-B_0 \frac{\partial v_x}{\partial x}\;, \label{eq_B1Dx}
\end{eqnarray}
with $\nu^{\prime}=4\nu/3$, where $\nu$ is the kinematic viscosity, and $p_T$ is the perturbed total pressure, given by
\begin{equation}
p_T=p_1+\frac{B_0B_y}{\mu_0}\;. \label{def_pT}
\end{equation}
The equation governing the final equilibrium state can be obtained using Eq. (\ref{eq_v1Dx}). At equilibrium, the time dependence disappears, and the velocity is zero, thus, the equilibrium must have constant total pressure:
\begin{equation}
\frac{\partial p_T}{\partial x}=0\;. \label{equilib1Dx}
\end{equation}
Combining Equations (\ref{eq_p1Dx}) and (\ref{eq_B1Dx}), we get the evolution of the total pressure as
\begin{equation}
\frac{1}{\rho_0}\frac{\partial p_T}{\partial t}=-(c^2_s+c^2_A) \frac{\partial v_x}{\partial x}\;, \label{eq_pT}
\end{equation}
where $c_s$ is the sound speed, defined above, and $c_A=B_0/\sqrt{\mu_0\rho_0}$ is the Afv\'en speed. By the appropiate combination of Equations (\ref{eq_v1Dx}) and (\ref{eq_pT}) we get the wave equation for fast magneto-acoustic waves for the total pressure:
\begin{equation}
\frac{\partial^2 p_T}{\partial t^2}=(c^2_s+c^2_A)\frac{\partial^2 p_T}{\partial x^2}+\nu^{\prime}\frac{\partial}{\partial t}\left(\frac{\partial^2 p_T}{\partial x^2}\right)\;. \label{wave_eq}
\end{equation}
Assuming that the total pressure can be considered as a continuous, periodic function, the solution of the last equation can be expressed as a superposition of plane waves, such as
\begin{equation}
p_T(x,t)=\mathcal{R}e\left(\sum_k\varphi_k{\rm e}^{{\rm i}(kx-\omega t)}\right)\;, \label{supwaves}
\end{equation}
where each wave number $k$ corresponds to a different oscillation mode and is associated with a complex frequency $\omega (k)$ given by the dispersion relation,
\begin{eqnarray*}
\omega^2+{\rm i}k^2\nu^{\prime}\omega-(c^2_s+c^2_A)k^2=0\;. \nonumber
\end{eqnarray*}
The dispersion relation has the solution $\omega=a-b{\rm i}$, where $a$ is the real frequency of the wave, and $b$ is the damping term:
\begin{eqnarray*}
a&=&\frac{k}{2}\sqrt{4(c^2_s+c^2_A)-k^2\nu^{\prime 2}}\;, \\
b&=&\frac{1}{2}k^2\nu^{\prime}\;.
\end{eqnarray*}
In order to have a harmonic mode, the wave number $k$ must satisfy $k^2\nu^{\prime 2}<4(c^2_s+c^2_A)$, and higher modes will be damped without any type of oscillation. On the other hand, notice that $\omega=0$ when $k=0$. The undamped mode $k=0$ corresponds to the constant Fourier coefficient in the expansion of Eq. (\ref{supwaves}), which does not change in time, and is given by Fourier analysis as the homogeneous redistribution of the initial total pressure to its average value. As $t\to\infty$, it is only this constant that remains, as all the other terms are proportional to ${\rm e}^{-bt}$. Hence, this homogeneous redistribution is exactly the constant perturbed total pressure that defines our final equilibrium state:
\begin{equation}
p_T(\infty)=\frac{1}{L_x}\int_x\!\left(p_1(x,0)+\frac{B_0B_y(x,0)}{\mu_0}\right)\,dx\;, \label{sol_pT}\\
\end{equation}
where $L_x$ is the length of the $x$-domain.

From the solution of Eq. (\ref{wave_eq}), and Eq. (\ref{eq_pT}), we obtain an expression for $v_1(x,t)$, which, after substitution into Equations (\ref{eq_rho1Dx}), (\ref{eq_p1Dx}) and (\ref{eq_B1Dx}), gives
\begin{eqnarray}
\frac{\partial \rho_1}{\partial t}&=&\frac{1}{c^2_s+c^2_A}\frac{\partial p_T}{\partial t}\;, \\
\frac{\partial p_1}{\partial t}&=&\frac{c^2_s}{c^2_s+c^2_A}\frac{\partial p_T}{\partial t}\;, \\
\frac{\partial B_y}{\partial t}&=&\frac{B_0}{\rho_0(c^2_s+c^2_A)}\frac{\partial p_T}{\partial t}\;.
\end{eqnarray}
Integrating now from $t=0$ to $t=\infty$, we obtain the perturbed quantities for the final equilibrium state, as functions of the perturbed total pressure, to be added to the background values:
 \begin{eqnarray}
\rho_1(x)&=&\rho_1(x,0)+\frac{1}{c^2_s+c^2_A}(p_T(\infty)-p_T(x,0))\;, \label{sol_rho} \\
p_1(x)&=& p_1(x,0)+\frac{c^2_s}{c^2_s+c^2_A}(p_T(\infty)-p_T(x,0))\;, \label{sol_p} \\
B_y(x)&=&B_y(x,0)+\frac{B_0}{\rho_0(c^2_s+c^2_A)}(p_T(\infty)-p_T(x,0))\;. \label{sol_B}
\end{eqnarray}

Equations (\ref{sol_rho}), (\ref{sol_p}) and (\ref{sol_B}) state that no matter how we set our initial disturbance, the final equilibrium distributions are completely determined by the initial and the final total pressures of the system, which are given by the solution of the wave equation. Note, that also the adiabatic equation for the linear regime given in Eq. (\ref{adiabatic}) is satisfied.


\subsection{Perturbation dependent on y} \label{sec:1Dy}

In the same way as above, we can get the solution for a perturbation varying only along the field. This time, we are dealing with a purely non-magnetic evolution, that will lead to a homogeneous redistribution of the plasma pressure all along the field lines. The equations to solve are the $y$-components of Equations (\ref{eq_rho1}), (\ref{eq_vy}) and (\ref{eq_p1}). The equilibrium is now given by
\begin{equation}
\frac{\partial p_1}{\partial y}=0\;, \label{equilib1Dy}
\end{equation}
and the solution for the perturbed quantities is
\begin{eqnarray}
p_1(\infty)&=&\frac{1}{L_y}\int_y\!p_1(y,0)\,dy\;,\\
\rho_1(y)&=&\rho_1(y,0)+\frac{1}{c_s^2}(p_1(\infty)-p_1(y,0))\;,
\end{eqnarray}
where $L_y$ is the length of the $y$-domain.


\subsection{Two-dimensional perturbation} \label{sec:2D}

In this section, we study the relaxation of a general individual two-dimensional perturbation within the linear regime. Once again, setting the velocities to zero, we get the equations governing the final 2D equilibrium:
\begin{eqnarray}
\frac{\partial}{\partial x}\left(p_1+\frac{B_0B_y}{\mu_0}\right)-\frac{B_0}{\mu_0}\frac{\partial B_x}{\partial y}&=&0\;, \label{equi_x}\\
\frac{\partial p_1}{\partial y}&=&0\;. \label{equi_y}
\end{eqnarray}

Equation (\ref{equi_y}) tells us that the final plasma pressure cannot depend on $y$, so the solution for the pressure must remain one-dimensional, as before. On the other hand, Eq. (\ref{equi_x}) does not have a direct interpretation, as both spatial derivatives are involved. The term $p_1+B_0B_y/\mu_0$ represents the perturbed total pressure from Sec. \ref{sec:1Dx}, and the new term $B_0B_x/\mu_0$ represents the magnetic tension due to the curvature of the field lines, which was zero in the 1D cases. As usual, when looking for a periodic solution, Fourier analysing makes life much simpler: Expressing our variables as functions of ${\rm e}^{{\rm i}(kx+ly-\omega t)}$, where each pair $(k,l)$ represents one single mode of oscillation in the global time evolution, Equations (\ref{equi_x}) and (\ref{equi_y}) can be rewriten as
\begin{eqnarray*}
k\,\left(p_1+\frac{B_0B_y}{\mu_0}\right)-l\,\frac{B_0B_x}{\mu_0}&=&0\;,\\
l\,p_1&=&0\;,
\end{eqnarray*}
where:
\begin{enumerate}[i.]
\item The mode $k=0,\;l=0$ represents the unperturbed background values.
\item For $k\neq0,\;l=0$, the equation of the equilibrium is
\begin{equation}
\frac{\partial}{\partial x}(p_1+B_0B_y)=0\;.
\end{equation}
These modes only depend on $x$, and represent the homogeneous redistribution of the total pressure studied in Sec. \ref{sec:1Dx}.
\item For $k=0,\;l\neq0$ we get
\begin{eqnarray}
\frac{\partial p_1}{\partial y}&=&0\;,\\
\frac{\partial B_x}{\partial y}&=&0\;.
\end{eqnarray}
These modes do not modify $B_y$, instead they simply remove both the vertical gradients of magnetic tension and plasma pressure as in Sec. \ref{sec:1Dy}. Each of them is treated individually, as they are not coupled in the equations.
\item Finally, for those modes with $k\neq0,\;l\neq0$, we get
$$kB_y-lB_x=0\;,$$
which can be combined with the solenoidal condition for the magnetic field, $\boldnabla\cdot{\bf B}=0$, or, within our fourier notation,
$$ kB_y+lB_x=0\;.$$
From these equations, we can conclude that, in the final equilibrium, the existance of a variation of $B_y$ in the $x$-direction is totally incompatible with a variation of $B_x$ in the $y$-direction. Hence, the modes with both wave numbers $k$ and $l$ non-zero may appear in the dynamical evolution, but {\it not} in the final equilibrium distributions.
\end{enumerate}

Therefore, with our uniform background magnetic field pointing straight in the vertical $y$-direction, the final equilibrium state is a combination of the background values ($k=0,\;l=0$), plus the vertical non-magnetic evolution to a state with plasma pressure that is constant along $y$, and/or the smoothing of the horizontal component of the magnetic field ($k=0$), plus the one-dimensional hydromagnetic evolution across the field lines ($l=0$). Note, that the perturbed $B_x$ in the vertical direction (i.e. curved magnetic field) is not coupled with either $p_1$ or $B_y$, so the final magnetic field remains as straight lines, and $B_x$ is not involved in the evolution of the total pressure of the system.

Hence, we calculate the analytical 2D equilibrium in two steps: Firstly, the non-magnetic evolution in the vertical direction, 
\begin{eqnarray}
p_1^*(x)&=&\frac{1}{L_y}\int_y\!p_1(x,y,0)\,dy\;, \label{midp1}\\
\rho_1^*(x,y)&=&\rho_1(x,y,0)+\frac{1}{c_s^2}(p_1^*(x)-p_1(x,y,0))\;, \label{midrho1}
\end{eqnarray}
and secondly, the hydromagnetic evolution in the horizontal direction, across the field,
\begin{eqnarray}
\rho_1(x,y)&=&\rho_1^*(x,y)+\frac{1}{c^2_s+c^2_A}(p_T(\infty)-p_T^*(x))\;,\\
p_1(x,y)&=&p_1^*(x)+\frac{c^2_s}{c^2_s+c^2_A}(p_T(\infty)-p_T^*(x))\;,\\
B_y(x,y)&=&B_y(x,0)+\frac{B_0}{\rho_0(c^2_s+c^2_A)}(p_T(\infty)-p_T^*(x))\;. \label{finalB1}
\end{eqnarray}
In Equations (\ref{midp1}) to (\ref{finalB1}) the quantities with a superscript represent the state after the vertical evolution, with $p_T^*(x)=p_1^*(x)+B_0B_y(x,0)/\mu_0$.

Looking back at the equations, we see a well known result from magnetohydrostatics, namely: In equilibrium, and in the absence of gravity, the plasma pressure must be constant along the field lines. The constant plasma pressure in the $y$-direction given by Eq. (\ref{equi_y}) is aligned with the straight magnetic configuration, with no magnetic tension, for the final equilibrium state.

When analyzing the validity of the results above for a non-ideal experiment, it is important to remember that Equations (\ref{midrho1}) to (\ref{finalB1}) come from the linear approximation, but Eq. (\ref{midp1}) does not. Hence, we expect our analytical calculations for the pressure to hold for much larger initial perturbations than the ones for the density. If the initial pressure disturbance is not small, but the linear expression for the plasma pressure is still valid, the adiabatic condition (i.e. conservation of entropy), which is more robust than the linear calculations, gives us a better approximation for the final equilibrium plasma density, calculated as
\begin{equation}
\rho(x,y,t\to\infty)=\left(\frac{p(x,y,t\to\infty)\;\rho^{\gamma}(x,y,t=0)}{p(x,y,t=0)}\right)^{1/\gamma}\;. \label{better_rho}
\end{equation} 


\section{Numerical Experiments}\label{sec:numexp}

To investigate the validity of the analytical results, we have used Lare2D, a staggered Lagrangian-remap code with user controlled viscosity, to solve the full MHD equations, with the resistivity set to zero \citep[for further details, see][]{Arber01}. The numerical domain is a square box with a uniform grid of $256\!\times\!256$ points. The background magnetic field is pointing in the vertical $y$-direction and all the perturbations depend on both $x$ and $y$. The top and bottom boundaries of the box are periodic, so that the field lines are not line-tied. The boundaries on the left and right sides are closed. The choice of either closed or periodic side boundaries makes no difference for our experiments. There is neither mass nor energy flowing across the side boundaries.

The numerical code uses the normalised MHD equations, where the normalised magnetic field, density and lengths,
\begin{eqnarray*}
x=L\hat{x}\;,\;\;\;y=L\hat{y}\;,\;\;\;{\bf B}=B_n\hat{\bf B}\;,\;\;\;\rho=\rho_n\hat{\rho}\;,
\end{eqnarray*}
imply that the normalising constants for pressure, internal energy and current density are
\begin{eqnarray*}
p_n=\frac{B_n^2}{\mu_0}\;,\;\;\;\epsilon_n=\frac{B_n^2}{\mu_0\rho_n}\;\;\;{\rm and}\;\;\;j_n=\frac{B_n}{\mu_0 L}\;.
\end{eqnarray*}
We have used here the subscript $n$ for the normalising constants, instead of $0$, to avoid confusion with the initial background values. The {\it hat} quantities are the dimensionless variables with which the code works. The expression for the plasma beta can be obtained from this normalisation as
\begin{equation}
\beta=\frac{2\hat{p}}{\hat{B}^2}=\frac{2\hat{\rho}\hat{\epsilon}(\gamma -1)}{\hat{B}^2}\;.
\end{equation}

\begin{figure}[ht]
\centering
\includegraphics[scale=0.45]{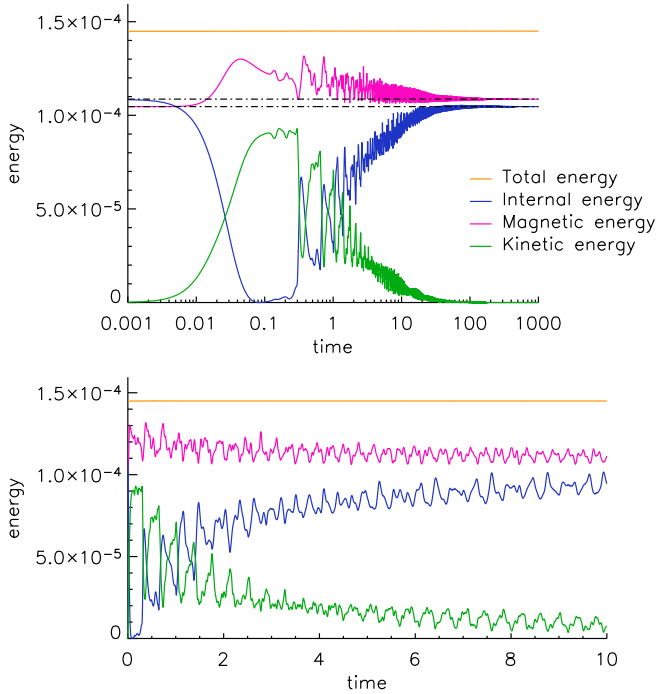}
\caption{Time evolution of the energies for the case where ${\rm max}(p_1)=p_0=0.1B_n^2/\mu_0$ and $\beta=0.2$, integrated over the whole 2D box. The four evolutions have been shifted in the vertical axis by substracting the constant values 0, 0.5, 0.15 and 0.65, respectively, for kinetic, magnetic, internal and total energy, but their relative amplitudes are not scaled. The final losses of internal energy are entirely balanced with a net increase of magnetic energy. The top figure is logarithmic in time and covers the whole relaxation. The bottom figure is linear in time and only covers the first part of the relaxation. From this graph, we can appreciate the complex oscillation periods, which are products of the sum of the different plane waves that drive the relaxation.}
\label{fig:energy}
\end{figure}

The initial disturbance consists of an internal energy and pressure enhancement, leaving the initial plasma density and magnetic field undisturbed. The perturbation is taken as a single two-dimensional Gaussian centered in the middle of the box:
\begin{equation}
\epsilon_1(x,0)=a\;{\rm exp}\left(-\frac{(x-b)^2}{2c^2}\right){\rm exp}\left(-\frac{(y-b)^2}{2c^2}\right)\;,
\end{equation}
with $b=0.5L$ and $c=0.05L$. As there is no initial perturbation of the magnetic field, the initial perturbed total pressure is just the perturbed plasma pressure, and the final analytical perturbed total pressure is the average value of that initial perturbed plasma pressure distribution, so that
\begin{eqnarray*}
p_T(\infty)-p_T^*(x)=\int_x\!\left(\int_y\!p_1(x,y,0)\,dy\right)\,dx\;-\;\int_y\!p_1(x,y,0)\,dy\;.
\end{eqnarray*}

For the experiment shown here, we have chosen a rather large perturbation, of the same order as the background value, i.e. $a=\epsilon_0$, for which one may {\it expect} linear theory {\it not} to be applicable. The background plasma beta is $0.2$.

Figure \ref{fig:energy} shows the time evolution of the various energies of the system, integrated over the whole box. Kinetic energy (in green) grows quickly from zero to its maximum value, and is subsequently damped to zero in the final equilibrium. A small fraction of the internal energy is converted into magnetic energy at the new equilibrium. For this particular set up, in which the perturbation has been introduced in the plasma pressure, it is the plasma that loses some of its initial internal energy, transfering it to the magnetic field. Note, that the amount of energy transfered is directly proportional to the magnitude of the perturbation. The total energy (i.e. the sum of the magnetic, kinetic and internal energies) is conserved to an accuracy of $\sim 10^{-7}$.

\begin{figure}[ht]
\centering
\includegraphics[scale=0.45]{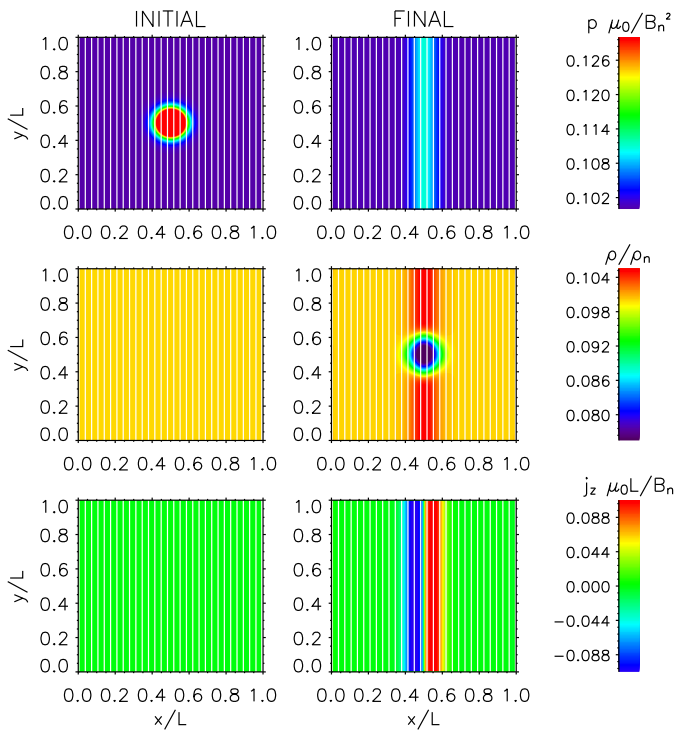}
\caption{Two-dimensional contour plots of plasma pressure (top), plasma density (center) and current density (bottom) in the initial state (left column) and final equilibrium (right column), for the same experiment as Fig. \ref{fig:energy}.}
\label{fig:2Dmaps}
\end{figure}

The 2D contour plots of the normalised plasma pressure, density and perpendicular current density for the initial and final equilibrium state are shown in Fig. \ref{fig:2Dmaps} for this case. The initial increase of plasma pressure creates a localized decrease in plasma density. The displacement of the magnetic field lines is hard to appreciate from the field lines themselves, but is clear from the non-zero perpendicular electric current density, $j_z$, that exists in the final equilibrium.

\begin{figure}[ht]
\centering
\includegraphics[scale=0.45]{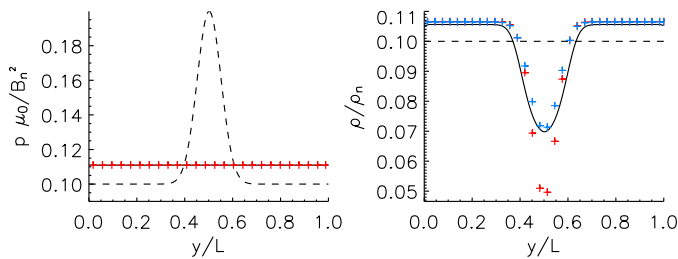}
\caption{Vertical cuts for the plasma pressure (left) and plasma density (right), for the same experiment as Fig. \ref{fig:energy} and Fig. \ref{fig:2Dmaps}. Initial perturbed state (dashed) is compared with the final equilibrium, as found by the full MHD numerical simulations (solid) and predicted by the linear analysis (red crosses). For the density predictions, the blue crosses represent predictions from the adiabatic condition given by Eq. (\ref{better_rho}).}
\label{fig:mhs2Dy}
\end{figure}

\begin{figure}[ht]
\centering
\includegraphics[scale=0.45]{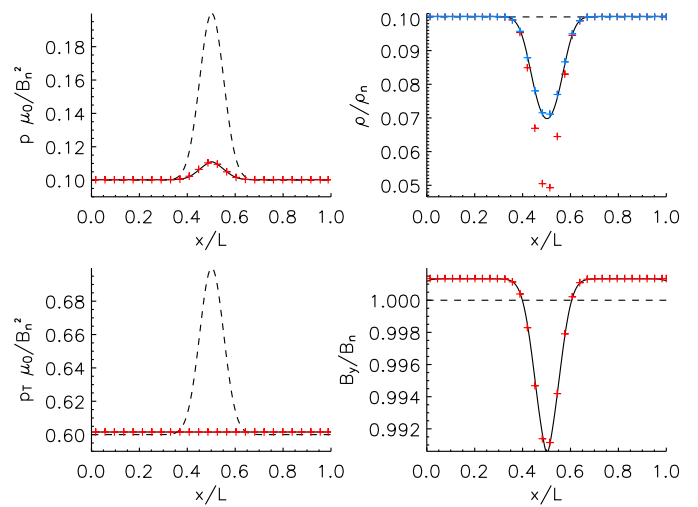}
\caption{Horizontal cuts for the plasma pressure (top left), plasma density (top right), total pressure (bottom left) and magnetic field strength (bottom left), for the same experiment as Figures \ref{fig:energy}, \ref{fig:2Dmaps} and \ref{fig:mhs2Dy}.}
\label{fig:mhs2Dx}
\end{figure}

Figures \ref{fig:mhs2Dy} and \ref{fig:mhs2Dx} show vertical cuts at $x=L/2$ and horizontal cuts at $y=L/2$, respectively, through the contour plots in Fig. \ref{fig:2Dmaps}, for the relevant quantities for the initial and final state. In these plots, a detailed comparison of the numerical and analytical solutions can be made. The analytical predictions for total pressure are very good, even though the amplitude of the perturbation is relatively large ($a=\epsilon_0$) and so, strictly speaking, our analytical approximation should not hold. However, the linear approximation for the plasma density (red crosses in Figures \ref{fig:mhs2Dy} and \ref{fig:mhs2Dx}) does not fit well. Instead, if we use Eq. (\ref{better_rho}), a much better fit for the plasma density is found (blue crosses in the plots), implying that the process is approximately adiabatic. Since the numerical experiments have been performed using a full MHD code that solves the non-linear equations, the process is not entirely adiabatic, but must have a finite amount of viscous heating that will become important as the initial perturbation is increased.

\section{Importance of non-linear effects} \label{sec:nonlin}

To study how non-linearity affects the results as the magnitude of the initial perturbation increases, we focus again on the total pressure. The total pressure of the final numerical equilibrium must be constant, whether the relaxation remains in the linear regime or not. On the other hand, the analytical definition of total pressure given by (\ref{def_pT}) is an approximation from the linear analysis, and will become less valid as the non-linear terms become more important. We perform a series of experiments for various plasma beta values in which the relative amplitude of the initial perturbation is changed from a very small value, well within the linear regime, to a large value way outside it. Using these experiments, we investigate how the final total pressure departs from the linear predictions for different background plasma beta values.

But first, we recall that the 2D relaxation may be separated into a vertical non-magnetic evolution (vertical redistribution of plasma pressure) and a horizontal evolution (horizontal redistribution of total pressure), in which the total pressure in the vertical case is {\it not} determined by the linear analysis. This suggests that, effectively, in order to find a significant error in the final total pressure for the 2D experiment, we will need very large values of the initial two-dimensional perturbation. Hence, the following experiments have been made for just a one-dimensional perturbation across the field lines. These results may be mapped onto those for our intital 2D perturbation, using the following:
\begin{eqnarray*}
\left(\frac{{\rm max}(p_1)}{p_0}\right)_{2D}&=&\frac{L}{\int_y\!{\rm exp}\left(-(y-b)^2/2c^2\right)\,dy}\;\left(\frac{{\rm max}(p_1)}{p_0}\right)_{1D}\;.
\end{eqnarray*}

\begin{figure}[ht]
\centering
\includegraphics[scale=0.37]{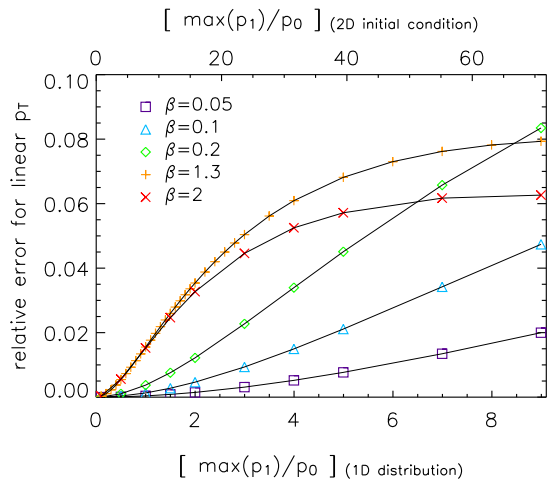}
\caption{Relative error in the linear prediction of the total pressure against the magnitude of the initial pressure perturbation, for five different values of the plasma beta. The slow growth rate of the error (non-linear effects) indicates the validity of the linear analysis studied in this paper.}
\label{fig:series}
\end{figure}

Figure \ref{fig:series} shows the relative error of the linear aproximation in both 1D and 2D for the total pressure, as a function of the amplitude of the initial perturbation, for five different values of the plasma beta ($\beta=0.05$, $\beta=0.1$, $\beta=0.2$, $\beta=1.3$ and $\beta=2$). The bottom $x$-axis shows the magnitude of the one-dimensional perturbation, and the top $x$-axis shows the magnitude of the initial two-dimensional perturbation before its vertical expansion. The error on the $y$-axis is calculated as the maximum difference between the linear prediction and the numerical results for the total pressure, 
\begin{eqnarray*}
\frac{{\rm max}(|p^{lin}_T-p^{num}_T|)}{p^{num}_T}\;,
\end{eqnarray*}
where $p^{num}_T$ is the final constant total pressure obtained from the numerical simulations, and, for our non-magnetic initial perturbation,
\begin{eqnarray*}
p^{lin}_T(x)&=&p_0+p_1(x,\infty)+\frac{B_0^2}{2\mu_0}+\frac{B_0B_y(x,\infty)}{\mu_0}\;,
\end{eqnarray*}
with $p_1(x,\infty)$ and $B_y(x,\infty)$ being the final perturbed plasma pressure and perturbed magnetic field from the numerical simulations.

As $\beta\to\infty$, we expect the relative error of the linear analysis to tend to zero, independently of the perturbation, as in this case, the magnetic effects dissapear, and the initial pressure perturbation completely redistributes to a well defined constant value in the whole box. On the other hand, if $\beta\ll 1$, then the magnetic field will dominate over the plasma contributions, and much larger values for the {\it relative} initial perturbation will be needed for strictly leaving the linear regime. These two behaviors can be seen in Fig. \ref{fig:series}, where the plots for large plasma-betas tend to a smaller error, while the plots for small plasma-betas take longer to reach significant errors, i.e. to escape from the linear regime. Furthermore, we must not forget that we are here only talking about the initial background plasma beta, so a large background beta combined with a large initial perturbation will make the final plasma beta even higher. Thus, ${\rm max}(p_1)/p_0\to\infty$ will imply $\beta\to\infty$ for the final equilibrium, so we expect the curves of the relative error of the linear analysis to turn back to zero as the initial perturbation is greatly increased. In terms of energy conservation, as the velocity is zero at the initial and final states, the integral over the whole domain of internal energy plus magnetic energy must be conserved: If $\beta\to\infty$, then the internal energy is much larger than the magnetic energy, and will just redistribute the plasma pressure, without transferring any energy into the magnetic field.

\begin{figure}[ht]
\centering
\includegraphics[scale=0.37]{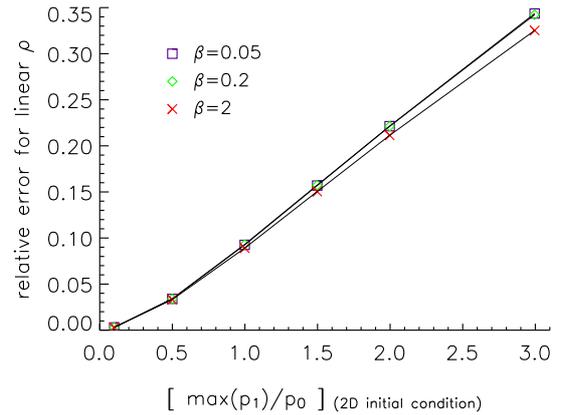}
\caption{Relative error of the density predicted assuming an adiabatic evolution, with Eq. \ref{better_rho}, against the magnitude of the 2D initial pressure perturbation, for three values of the plasma beta. Note, that the $x$-axis in this plot is to be compared with the top $x$-axis in Fig. \ref{fig:series}.}
\label{fig:series_rho}
\end{figure}

On the contrary, the final plasma density is entirely determined by the linear analysis, in both the vertical and the horizontal evolutions along and across the field lines, or in a better approximation, by the adiabatic condition. Hence, the non-linear effects for the plasma density will grow much quicker, as shown in Fig. \ref{fig:series_rho}. These last numerical experiments have been made for the original two-dimensional Gaussian perturbation. The error on the $y$-axis is given by
\begin{eqnarray*}
\frac{{\rm max}(\rho^{ad}-\rho^{num})}{\rho^{num}}\;,
\end{eqnarray*}
where $\rho^{ad}$ is the plasma density given by Eq. (\ref{better_rho}), and $\rho^{num}$ is the final density obtained with the numerical experiments.

The relative error in the plasma density is considerably bigger than the relative error in pressure, and so, for only a small change in $p_1/p_0$ in the 2D case, we find a large error in $\rho$. As this error quickly reaches significant values, the plasma beta plays much less of a role for the non-linear effects in the plasma density than in the above total pressure.

\section{Summary and Conclusions} \label{sec:summary}

We have presented analytical and numerical calculations for the 2D magnetohydrodynamic relaxation of an untwisted perturbed magnetic system embedded in a plasma with beta of any size, and find a final equlibrium state which differs substantially from the initial background configuration. The equilibrium reached is a non force-free state in which the plasma pressure gradients are balanced by the magnetic Lorentz force. For a set of specified boundaries, all the hydromagnetic quantities are fully determined by the initial perturbed state.

The initial disturbance evolves into the final relaxed state by different families of magnetoacoustic waves, dissipated via viscous damping. Fast magnetoacoustic waves propagate across the field lines in the horizontal direction, slow magnetoacoustic waves redistribute the thermodynamic quantities along the field in the vertical direction, and an extra contribution of slow magnetoacoustic waves propagates along the magnetic field lines introducing a magnetic tension term. Nevertheless, these last slow magnetoacoustic waves dissipate the magnetic tension in such a way that it is totally unimportant when determining the final equilibrium distributions. The vertical redistribution of the plasma pressure to a homogeneous value demands the magnetic tension to dissapear completely, so both the plasma pressure and total pressure are one-dimensional at the end of the process.

Within the linear regime, the final distributions are completely independent of the viscosity, even though it is required to permit the relaxation to ocurr, as it is the only damping mechanism of the waves. An increase in the viscosity enhances the diffusive term in the wave equation, and so, accelerates the process, but the final distribution is not modified. Instead, in the final equilibrium, all the quantities are simply determined by the behaviour of the final equilibrium total pressure, involving plasma and magnetic effects. Hence, the final equilibrium states for plasma pressure and magnetic field do not differ if the initial perturbation is of the density, temperature or internal energy.

Finally, by investigating the linear regime, we have been able to make analytical predictions for the final MHS equilibrium, even when the regime is far from linear. The linear predictions remain remarkably valid even outside the linear regime, as the growth rate of the non-ideal effects is very small, compared to the initial perturbations.

In this paper, the introduction of plasma effects in the relaxation of hydromagnetic systems have been studied in simple schemes, producing a series of analytical predictions which have been confirmed by our numerical results. By starting with a uniform magnetic field, we have reached a state in which the magnetic field itself remains almost uniform, but where some current density has been built. Also, even in this simple configuration, a non-neglectible amount of energy has been transferred from the plasma to the magnetic field. These implications of energy transfer during the relaxation process indicate that this process will be of importance in the solar corona, specially in the study of magnetic null points and their surroundings. Null points have been found to have a reasonable population density in the Solar Corona by \citet{Longcope09}. In further studies, we will consider more complex scenarios, introducing two-dimensional null points and starting to consider the implications of non-zero plasma betas in three-dimensional magnetic environments.


\bibliographystyle{aa}
\bibliography{13902}


\end{document}